\documentclass{PoS}

\usepackage{mathtools}
\usepackage{subfigure}

\title{Lightening-like interactions in nuclear collisions at CERN large hadron collider}

\ShortTitle{Lightening-like interactions in nuclear collisions}

\author{{Khaled Abdel-Waged}\\
        Umm Al-Qura University, Faculty of Applied Science, Physics Department, P.O. Box (715), Makkah 21955,
 Saudi Arabia\\
        E-mail: \email{kamabdellatif@uqu.edu.sa}}

\author{{Nuha Felemban}\\
       Umm Al-Qura University, Faculty of Applied Science, Physics Department, P.O. Box (715), Makkah 21955,
Saudi Arabia\\
     E-mail: \email{nafelemban@uqu.edu.sa}}

\abstract{A simple basic model for describing proton-nucleus $(p+A)$ and nucleus-nucleus $(A+A)$ collisions has been the intra-nuclear cascade model, where the interactions are simulated by a sequence of binary nucleon-nucleon $(NN)$ collisions. This model helped to establish many scientific concepts and also creates the foundation for more modern simulation codes, especially at low and intermediate energies. In this paper, we present a new Monte Carlo model for $p+A$ and $A+A$ collisions at high CERN Large Hadron collider energies. The model implements HIJING code  with a collective cascade recipe, that induces lightening-like effect in a large nucleus. A single collision (lightening) event is shown to be a complex process: A primary interacting nucleon passes its energy to the surrounding nucleons in a large nucleus. This new simulation code is shown to be good to reproduce the Large Hadron collider (LHC) data, especially the charged particle pseudorapidity density in $p+Pb$ and $Pb+Pb$ collisions at LHC energies. }

\begin{document}

\section{Introduction}
Microscopic transport codes [1-6] are a very powerful tool to simulate proton-nucleus $(p+A)$ and nucleus-nucleus $(A+A)$ collisions at Relativistic Heavy Ion Collider (RHIC) and Large Hadron Collider (LHC) energies. All of these codes are based on intra-nuclear cascade models, where $p+A(A+A)$-interactions are treated as a set of binary nucleon-nucleon $(NN)$ collisions. At $NN$-center of mass energies ($\sqrt{s_{NN}}$) greater than 100 GeV, multiple mini-jet production combined with soft interactions are the main mechanisms that provide an adequate description of $p+p(\bar{p})$ RHIC and LHC data. However, for $p+A$ and $A+A$ collisions, where nuclear effects play an important role, none of these codes was able to describe all LHC data simultaneously, especially the pseudorapidity density and transverse momentum distributions of charged particles in $p+Pb$ and $Pb+Pb$  collisions  at  $\sqrt{s_{NN}}=5.02$ and $2.76$ TeV, respectively.

In this paper we present a new Monte Carlo model for $p+A$ and $A+A$ collisions at high CERN LHC energies. The model supplements HIJING code \cite{r2} with updated parton distribution functions \cite{r7} and a collective cascade recipe \cite{r8,r9}, that induces lightening-like effect in a nucleus. In Sec.2 the basic principles of  the code is discussed. In Secs.3 and 4, the effect  is demonstrated by comparing the model predictions with the recent measured ALICE data of $p+Pb$ and $Pb+Pb$  collisions [10-13]. Finally, we conclude this work in Sec.5.

\section{The model}
In this section,  HIJING 1.0 code \cite{r2} with an updated parton distribution functions \cite{r7} (in short, improved HIJING (ImHIJING)) is supplemented with a collective cascade recipe. Aiming to establish a simple standard model, we have chosen the standard type of HIJING model \cite{r2}.

\begin{figure}[htbp]
\begin{center}
\includegraphics[scale=0.5]{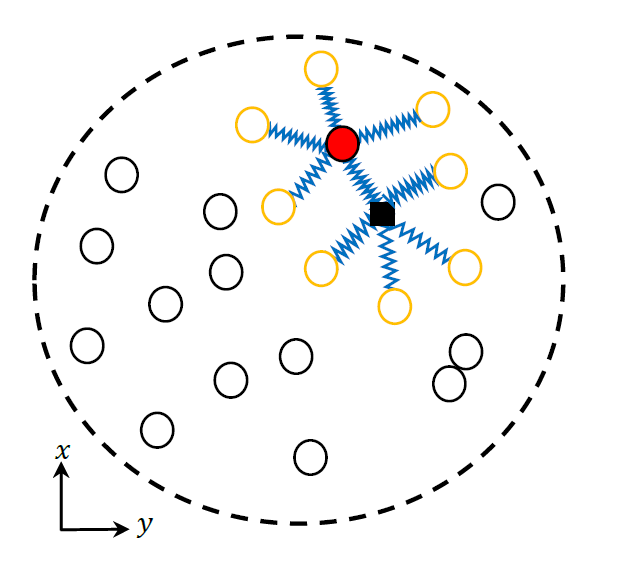}
\caption{(COLOR ONLINE)
A schematic representation of lightening-like effect for a single $p+A$ collision in the impact parameter plane. All nucleons are shown as open circles, a primary interacting nucleon is marked by closed circle, the set of individual reggeon exchanges by wavy lines, and the square point is the reggeon interaction vertex.}
\label{f1}
\end{center}
\end{figure}

HIJING model \cite{r2,r3} describes  $A+A$ interactions as a set of binary $NN$- collisions. At a given impact parameter $(\vec{b})$ and given center of mass energy $(\sqrt{s})$,  $NN$ scatterings are handled by the eikonal formalism. Particles produced from two colliding nucleons at high energies ($\sqrt{s_{NN}}>4$ GeV) are described by a hard and a soft components. The hard component involves processes in which mini-jets are produced with transverse momentum $p_T$ larger than a transverse momentum cut off $p_0$. The inclusive cross section $\sigma_{jet}$  of the mini-jets is described by perturbative QCD, which depends on the parton-parton cross section $\sigma_{ab}$, parton distribution function $f_{a(b)} (x_{a(b)}, Q^2 )$  and $p_0$, where $x_{a(b)}$ is the light cone fraction momentum of parton $a$($b$). The kinematics of the jets and the associated initial and final state radiation are simulated by PYTHIA model \cite{r14}. On the other hand,  the soft component $(p_T <p_0)$, characterized by a soft cross section $\sigma_{soft}$, treats non-perturbative processes and is modelled by the formation and fragmentation of strings,  along the lines of FRITIOF \cite{r15} and DPM \cite{r1,r16} models.

ImHIJING  is an improved version of  HIJING1.383  in which the old Duke-Owen (DO1984) \cite{r17}  parameterizations of parton distribution functions (PDFs)  are replaced by a more modern sets of Martin-Stirling-Throne-Watt  (MSTW2009) PDFs \cite{r7}. Compared to DO 1984 (and Gluck-Reya Vogt (GRV1995) parameterizations of HIJING2.0 \cite{r18}), the MSTW2009 include global fits to a larger number of  data sets, which includes both old and new types of data. The old data are improved in their precision and kinematic range. The new data include the most precise data of inclusive jet production from both HERA and Run II at the Tevatron from CDF  \cite{r19} and D{\O} \cite{r20,r21}, that goes to larger jet $p_T$ values. These data are important as it constrains the gluon (and quark) distributions in the domain $0.01\leq x \leq0.5$ \cite{r7}.

Using the MSTW2009 tabulated form of PDFs and following the same procedure as in HIJING 2.0 \cite{r3}, the two free parameters of the model $p_0$ and $\sigma_{soft}$  are taken as energy dependent and chosen to fit $p+p(\bar{p})$ total cross sections and the pseudorapidity density of charged particles $( dN_{ch}/d\eta)$ at mid-pseudorapidity.  With tuned $p_0 (\sqrt{s})$  and  $\sigma_{soft}  (\sqrt{s})$, ImHIJING is found to give  the best description of $dN_{ch}/d\eta$ , the multiplicity distributions of charged particles and transverse momentum spectra in non-single diffractive $p+p$ collisions at LHC energies, within the pseudorapidity interval $|\eta|<2.4$ \cite{r22}.

For high energy heavy-ion collisions, HIJING implements nuclear effects via nuclear modification of the parton distributions functions (so called parton shadowing) \cite{r2,r3}. It is assumed that the parton distributions in a nucleus (with mass number A), $f_{a/A} (x_a,Q^2)$ are factorizable into parton distributions in a nucleon $f_{a/N}(x_a, Q^2)$  and the parton, \emph{a}, shadowing factor $R_{a/A} (x_a)$ \cite{r2,r3},
\begin{equation}\label{eq1}
f_{a/A}(x_{a}, Q^{2})=R_{a/A} (x_a)f_{a/N}(x_a, Q^2)
\end{equation}
In default HIJING1.383, the shadowing effect for quarks $(q)$ and gluons $(g)$ is taken the same, and the  impact-parameter dependent but $Q^2$-independent parameterizations is given by \cite{r2}
\begin{equation}\label{eq2}
\begin{split}
R_{a/A}(x_a, Q^2)&=\frac{f_{a/A}(x_a, Q^2)}{f_{a/N}(x_a, Q^2)}\\
&=1+1.19 \log^{1/6}⁡A[x_a^3-1.2 x_a^2+0.21x_a]\\
&-s_{q(g)}\frac{4}{3}\sqrt{1-r_{ij}^2/R_A^2}(A^{1/3}-1)\times[1-\frac{1.08}{A+1}\sqrt{x_a}]e^{-x_a^2/0.01}
\end{split}
\end{equation}
where $r_{ij}^2=\sqrt{(b_x+x_i-x_j)^2+(b_y+y_i-y_j)^2}$ is the transverse distance of the interacting nucleon pair (\emph{i} and \emph{j}) from its nucleus center, $b_{x(y)}$ and $x_{i(j)}$, $y_{i(j)}$ are the components of the impact parameter vector and the coordinates of the pair. The quark/gluon shadowing parameter $s_{q(g)}$  is constrained by the experimental data on the centrality dependence of the pseudorapidity density per participant pair.

The lightening-like effect in $p+A$ and $A+A$-collisions is treated in ImHIJING by utilizing a collective cascade recipe \cite{r8,r9} as follows.
\begin{enumerate}
  \item The primary interacting nucleons of the projectile (A) and target (B) nuclei are determined by means of the eikonal formalism as implemented in HIJING.
  \item Target and projectile spectator nucleons (those nucleons which have not been involved in the interaction) are then followed. If the $i^{th}$ spectator nucleon is at an impact distance $r_{ij}$ from the $j^{th}$ primary interacting nucleon, then it is considered to be participant of the collision with the probability $\varphi=C \exp(-r_{ij}^2/r_c^2) $,  where $r_c=1.2$ fm  is the mean interaction radius and C is a free parameter of the model, that determines the strength of secondary interactions. Such a nucleon can involve another spectator nucleon and so on. Note that in the case of  $C=0$, $\varphi$ reduces to the eikonal case, no secondary interactions.
  \item If the number of newly involved nucleons are not zero, then step (2) is repeated. This allows one to include 3,4,5... nucleons in the interaction, which  mimics  lightening-like interactions in nuclear collisions (see Fig.\ref{f1}).

  \item The $i^{th} (j^{th})$ wounded nucleon (the one which experiences a primary and/or a secondary interaction(s)) of the projectile (target) is ascribed a momentum  $\{{x'}_i^+ ({y'}_j^-) ,{p'}_{Ti} ({q'}_{Tj})\}$ distributed according to the law:

\begin{equation}\label{eq3}
   P({x'}_i^+,{p'}_{Ti})\varpropto\prod_{i=1}^{N_A}\exp(\frac{-{p'}_{Ti}^2}{<p_T^2>})  \exp(-\frac{({x'}_i^+-1/N_A )^2}{d^2})
   \end{equation}

 under the constraints $\sum_{i=1}^{N_A}{p'}_{Ti}=0 $ and  $\sum_{i=1}^{N_A}{x'}_i^+=1 $, where $N_{A(B)}$ is the number of wounded nucleons from the projectile(target). The values of \emph{d} and   $<p_T^2>$ are chosen as  0.5 and 0.5 (GeV/c)$^2$, which are fixed from the analysis of $dN_{ch}/d\eta$ in $p+Pb$ collisions at  $\sqrt{s_{NN}}=5.02$ TeV \cite{r22}.
 \item 	The final momentum $\{{x'}_i^+,{p'}_{Ti} \}$ and $\{{y'}_j^-,{q'}_{Tj} \}$ of the $i^{th} (j^{th})$ wounded nucleon is obtained in terms of \cite{r23}
   \begin{equation}\label{eq4}
      {p'}_{zi}=({W'}_A^+{x'}_i^+-\frac{{m'}_{Ti}^2}{{x'}_i^+ {W'}_A^+ })/2,
   \end{equation}
     \begin{equation}\label{eq5}
      {q'}_{zj}=-({W'}_B^- {y'}_j^--\frac{{\mu'}_{Tj}^2}{{y'}_j^- {W'}_B^- })/2.
     \end{equation}
 where ${m'}_{Ti}^2=m_i^2+{p'}_{Ti}^2, {\mu'}_{Tj}^2=\mu_j^2+{q'}_{Tj}^2$, and $m_i (\mu_j)$ is the mass of the $i^{th} (j^{th})$ wounded nucleon from A(B). ${W'}_A^+$ and ${W'}_B^-$ are constrained from energy-momentum conservation and given by \cite{r23}
 \begin{equation}\label{eq6}
  {W'}_A^+=\frac{W_0^- W_0^++\alpha-\beta+\sqrt{\Delta}}{2W_0^-},
\end{equation}
\begin{equation}\label{eq7}
  {W'}_B^+=\frac{W_0^- W_0^+-\alpha+\beta+\sqrt{\Delta}}{2W_0^+},
\end{equation}
where
\begin{equation*}
  W_0^+=(E_A^0+E_B^0 )+(p_{zA}^0+q_{zB}^0 ),
\end{equation*}
\begin{equation*}
 W_0^-=(E_A^0+E_B^0 )-(q_{zA}^0+q_{zB}^0 ),
\end{equation*}
 \begin{equation*}
\alpha=\sum_{i=1}^{N_A}\frac{{m'}_{Ti}^2}{{x'}_i^+},  \beta=\sum_{j=1}^{N_B} \frac{{\mu'}_{Tj}^2}{{y'}_j^-}
 \end{equation*}
 and
\begin{equation*}
\Delta=(W_0^- W_0^+ )^2+\alpha^2+\beta^2-2W_0^- W_0^+ \alpha-2W_0^- W_0^+ \beta-2\alpha \beta.
\end{equation*}
\end{enumerate}
In the above equations, the  initial total energy and momentum of nucleus A are defined by $E_A^0=\sum_{i=1}^{N_A}E_i$   and $p_{zA}^0=\sum_{i=1}^{N_A} p_{zi}$, respectively, and those of nucleus B are $E_B^0=\sum_{j=1}^{N_B} E_j$ and $q_{zB}^0=\sum_{j=1}^{N_B}q_{zj}$, where $E_i (E_j)$ and $p_{zi} (q_{zj})$ are the initial energy and longitudinal momentum of the $i^{th}$ $(j^{th})$ wounded nucleon, respectively.

In what follows, we denote the improvements established using the collective cascade recipe in ImHIJING as ImHIJING/Cas. In order to simulate lightening-like effect, the recipe is calculated with full cascading, $C=1$. In all calculations, the default HIJING1.383 parameters are selected and no adjustments are attempted.

\section{ALICE Proton-Lead results}

In this section we analyse the ALICE measurements \cite{r10,r11} of pseudorapidity density and transverse momentum spectra of charged particles in $p+Pb$ collisions at $\sqrt{s_{NN}}=5.02$ TeV. Because the charged particles spectra in $p+Pb$  collisions are measured in minimum bias (MB), we generate $2\times10^6$ events for a range of impact parameters from 0 to $2R_A$, i.e.,  20000 events are generated in equally spaced 100 impact parameter interval. As was done for experimental data, the analysis of  HIJING1.0 and ImHIJING (with and without lightening-like effect) generated events exclude single diffractive (SD) collisions.

\begin{figure}[htbp]
\begin{center}
\includegraphics[scale=0.5]{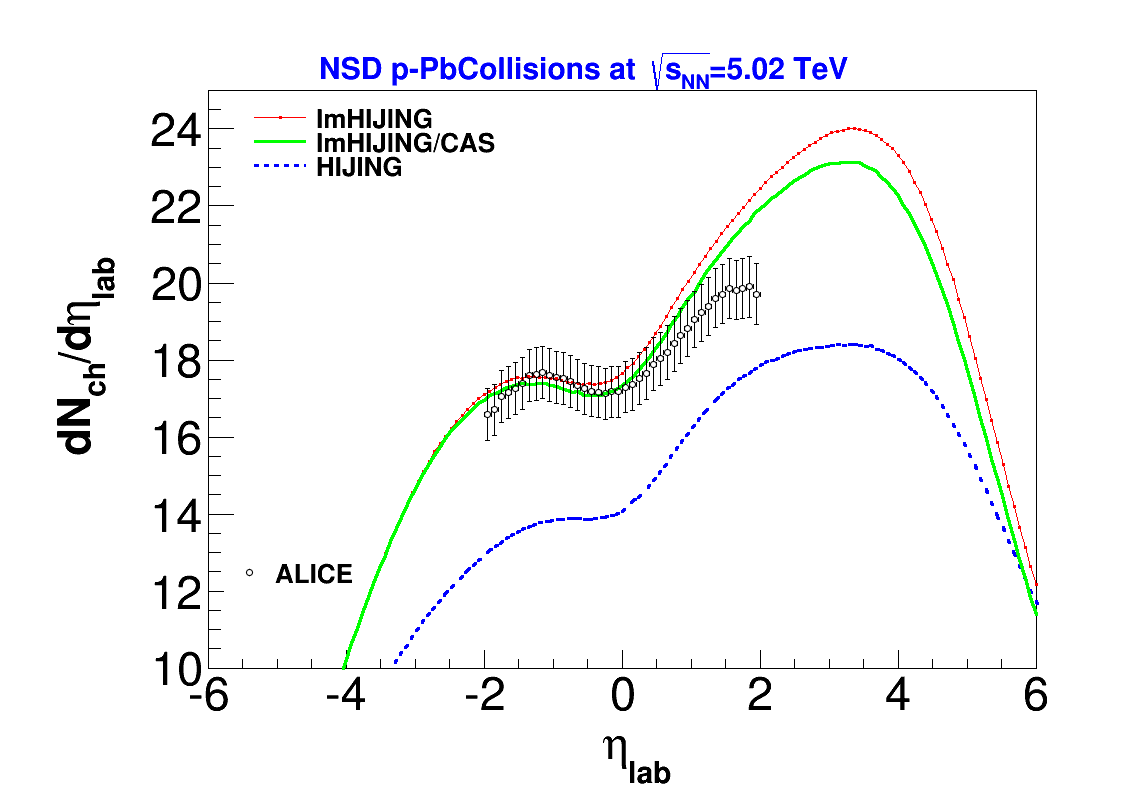}
\caption{(COLOR ONLINE)
Pseudorapidity density (in the laboratory system) of charged particles in NSD $p+Pb$ collisions at $\sqrt{s_{NN}}=5.02$ TeV, from ALICE experiment \cite{r10}, as compared to HIJING 1.0 and ImHIJING(/Cas) results.}
\label{f2}
\end{center}
\end{figure}

In Fig.\ref{f2}, we compare both the ImHIJING(/CAS) and HIJING1.0 results with the global observables of charged particle density in the laboratory system ($dN_{ch}/d\eta_{lab}$) for NSD $p+Pb$ collisions at $\sqrt{s_{NN}}=5.02$ TeV. The cut off parameter is  fixed at $p_0=5.4$ GeV/c. The HIJING1.0 calculations are with a constant cut off parameter  $p_0=2$ GeV/c and parton shadowing parameter of $s_{q(g)}=0.1$. As was done for the ALICE experiment \cite{r10}, both ImHIJING(/Cas) and HIJING 1.0 calculations assume that  the proton is moving in the negative $z$-direction at 4 TeV energy, while $Pb$ ion is moving in the positive $z$-direction at $82\times4$ TeV energy.

 It is interesting to note that the lightening-like effect increases when going from  proton $(\eta_{lab}\approx-1.4 0)$ to $Pb$ $(\eta_{lab}\approx3.4)$ peak regions. The ImHIJING/CAS results (thick line) lead to a better agreement with the data and can predict the forward-backward asymmetry shape of $dN_{ch}/d\eta_{lab}$. Fig.\ref{f2} also shows that the default HIJING1.0 describes the trend seen in the data, although it seems that, with the soft  shadowing parameter  $s_{q(g)}=0.1$, the model underpredicts the data.

\begin{figure}[htbp]
\begin{center}
\includegraphics[scale=0.5]{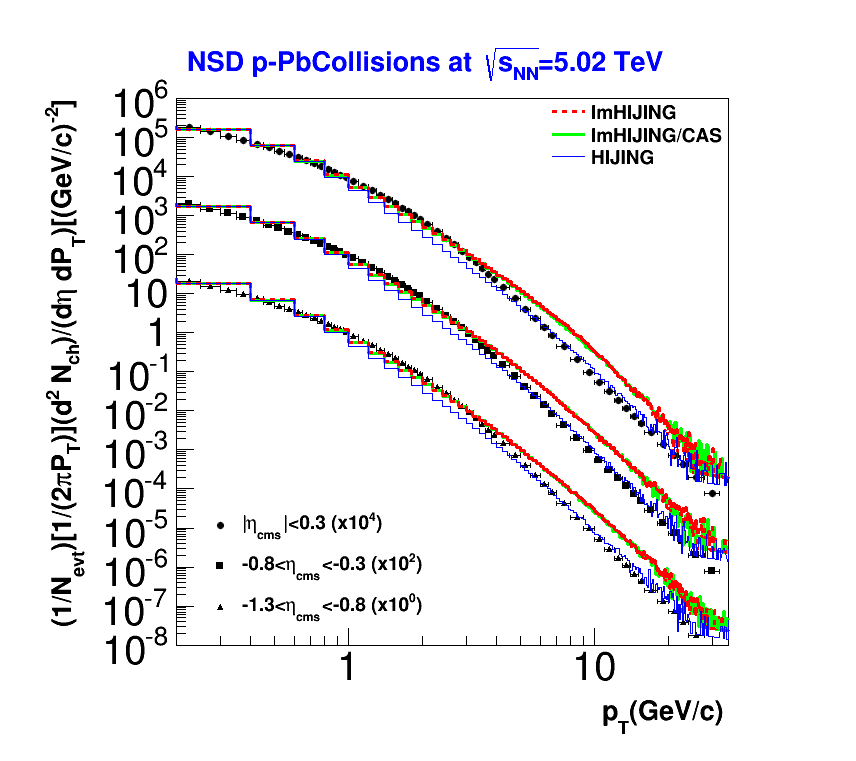}
\caption{(COLOR ONLINE)
Transverse momentum distributions of charged particles at different pseudorapidity intervals in NSD $p+Pb$ collisions at $\sqrt{s_{NN}}=5.02$ TeV, from ALICE experiment \cite{r11}, as compared to HIJING 1.0 and ImHIJING(/Cas) results. In order to avoid superposition of curves, both histograms and the data are multiplied by the indicated values.}
\label{f3}
\end{center}
\end{figure}

In Fig. \ref{f3} we examine the $p_T$-spectra of charged particles in the central $(|\eta_{cm} |<0.3)$ and backward $(-0.8<\eta_{cm}<-0.3 $ and $-1.3<\eta_{cm}<-0.8)$ pseudorapidity ranges. It is pointed out in Ref. \cite{r11} that, at $p_T\geq2$ GeV/c, the measured $p_T$-spectrum of charged particles in $p+Pb$ collisions is similar to those in an interpolated  $p+p$ reference spectra at $|\eta_{cm} |<0.3$, obtained by scaling data measured at $\sqrt{s_{NN}}=2.76$ and 7 TeV.  As one can see, the ImHIJING calculations nicely predicts the $p_T$-spectra at $p_T<6$ GeV/c. However, starting from $p_T\cong6$ GeV, the ImHIJING(/Cas) overestimates the measured  $p+Pb$ spectra.  This may imply that the strong suppression observed in $p+Pb$  collisions at $p_T>6$ GeV/c is not due to initial state effects but rather to final state interactions, which is not implemented for the current HIJING version.

In the region of low $p_T$ spectra ($p_T<6$ GeV/c) one thus can quantify lightening-like effects in  $p+Pb$ collisions at the specified pseudorapidity intervals. However, Fig.\ref{f3} demonstrates that there is almost no difference between the ImHIJING and ImHIJING/CAS results for the whole  $p_T$-spectra in all pseudorapidity ranges.

 On the other hand, calculations with default HIJING 1.0 are a good fit for both low ($p_T<1$ GeV/c)  and high ($p_T>6$ GeV/c) $p_T$-charged particles in $p+Pb$ collisions (see Fig.\ref{f3}). But we recall that the HIJING 1.0 calculated  $dN_{ch}/d\eta_{lab}$ yield already underestimates the measured distribution. This indicates that the HIJING 1.0 calculations are not consistent with the LHC data of charged particles in $p+Pb$ collisions. It should also be noted that calculations with DPMJET code \cite{r24}, which describes well the measured $dN_{ch}/d\eta_{lab}$\cite{r10}, overestimate the spectra by up to 22\% for  $p_T<0.7$ GeV/c and underestimate them by up to 50\% for  $p_T>0.7$ GeV/c\cite{r11}.

\section{ALICE Lead-Lead results}

In this section, we display the predictions of the  ImHIJING code (with and without lightening-like effect) along with the recent measurements of ALICE on the pseudorapidity density $(dN_{ch}/d\eta)$ and transverse momentum $(p_T)$  distributions of charged particles as a function of collision centrality in $Pb+Pb$ \cite{r12,r13} collisions at $\sqrt{s_{NN}}=2.76$ TeV. Because the LHC are measured in minimum bias (MB), we generate $7\times10^5$ events for a range of impact parameters from 0 to 2 $R_A$, i.e.,  7000 events are generated in equally spaced 100 impact parameter interval The ImHIJING(/Cas) calculations are performed with a constant cut off parameter  $p_0=4.5$ GeV/c  and an impact parameter dependence of parton shadowing. We use for the different centrality classes, the range of impact parameter provided by ALICE and not the one extracted from HIJING \cite{r25}.

\begin{figure}[htbp]
\begin{center}
\includegraphics[scale=0.5]{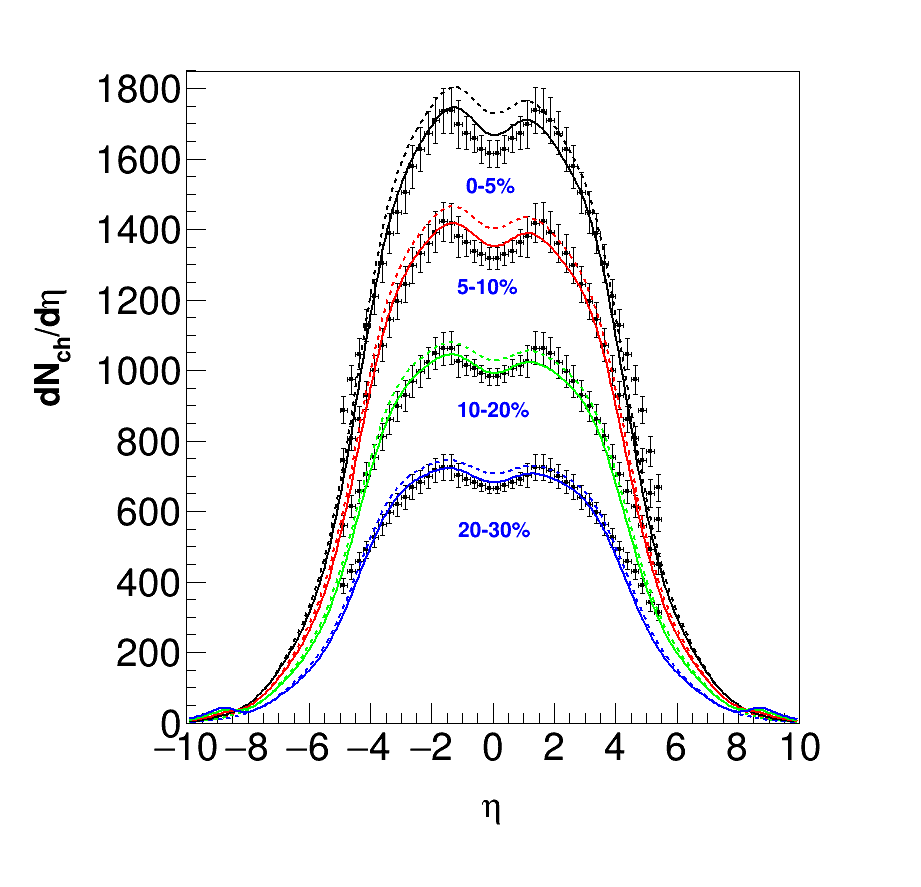}
\caption{(COLOR ONLINE)
Pseudorapidity density of charged particles in $Pb+Pb$ collisions at $\sqrt{s_{NN}}=2.76$ for the 30\% centrality intervals. The solid and short dashed lines denote the ImHIJING/Cas and ImHIJING calculations, respectively. The solid points with error bars denote ALICE data \cite{r12}.}
\label{f4}
\end{center}
\end{figure}

As shown in Fig.\ref{f4}, a reasonable description of the measured dependence of the charged particle pseudorapidity density on collision centrality is achieved with a form of $s_{q(g)}$ that depends on impact parameter

\begin{equation}\label{eq8}
s_{q(g)}(c)=s_{q(g)}+0.12 c^3-0.14c^2-0.0003c,
\end{equation}
with $s_{q(g)}=0.1$ and $c=\pi b^{2}/ \sigma_{in}$ is the centrality. It should also be noted that the estimated value of $s_{q(g)}$ in ImHIJING(/Cas)  is smaller than HIJING 2.0 (A MultiPhase Transport AMPT \cite{r26}) estimate of   $s_{q(g)}=0.2-0.23 (s_{q(g)}=0.16-0.17) \cite{r3,r27}$ which indicates the importance of using the most precise MSTW2009 tabulated form of PDFs, that constrain the gluon (and quark) distributions in the domain $0.01\leq x\leq0.5$, and lightening-like effect.

\begin{figure}[htbp]
\begin{center}
\includegraphics[scale=0.4]{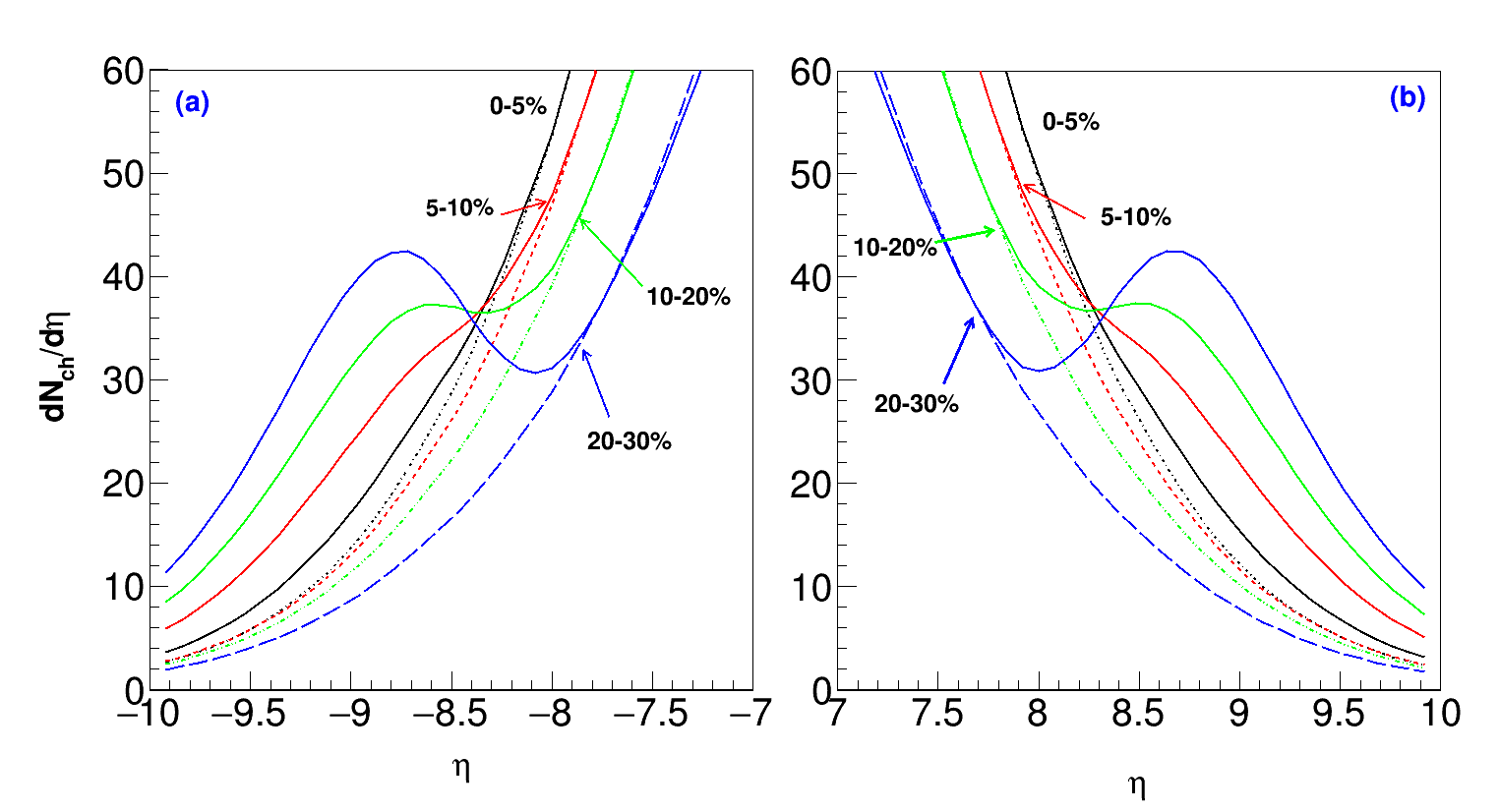}
\caption{(COLOR ONLINE)
Same as Fig. [4], but with a closeup of the pseudorapidity regions where the lightening-like effect is more pronounced.}
\label{f5}
\end{center}
\end{figure}

In Fig.\ref{f4}, we compare both the ImHIJING and ImHIJING/Cas results with the LHC data of $dN_{ch}/d\eta$ per  centrality class (from 0-5\% to 20-30\%) for the reaction under study. The two calculations differ only in the lightening-like effect, the impact parameter dependence of quark/gluon shadowing is identical in both cases. As one can see, the lightening-like effect starts to increase from the projectile/target pseudorapidity region (at $|\eta|\approx8$) reaching a peak  at $|\eta|\approx 1.5$ and then followed by a small fall-off to the center $(\eta=0)$. It is interesting to note that the largest effect develops as one moves toward more peripheral collisions  $(|\eta|>8)$, see Fig.\ref{f5}, where two peaks are built up and are enhanced with decreasing centrality. This indicates that the lightening-like effect induces nuclear modification of nucleons that are involved in the primary interactions. More specifically, nucleons taking part in the primary interactions suffer energy loss due to cascading with other non-interacting ones, that in turn influences the description of  $dN_{ch}/d\eta$ as a function of collision centrality.

 ImHIJING yields a similar shape to the measured results, but overestimates the level at mid-pseudorapidity $(|\eta|\leq2)$ with increasing centrality (dashed dotted lines). Attempts to tune ImHIJING by adjusting $s_{q(g)}(b)$ result in an underprediction of the centrality dependence of   $dN_{ch}/d\eta $ at  $|\eta|>4$ for ALICE data (not shown here). On the other hand, calculations with ImHIJING/Cas are a good fit for the whole centrality dependence of $dN_{ch}/d\eta$ (solid lines).

 It should be pointed out that, the UrQMD \cite{r5}, a Color Glass Condensate-type \cite{r28} and AMPT \cite{r26} (which uses HIJING2.0 as initial conditions) models, fail to reproduce the overall level and shape of ALICE data \cite{r12}. This may imply that the inclusion of lightening-like effect in ImHIJING is more consistent with LHC data.

\begin{figure}[htbp]
\begin{center}
\includegraphics[scale=0.5]{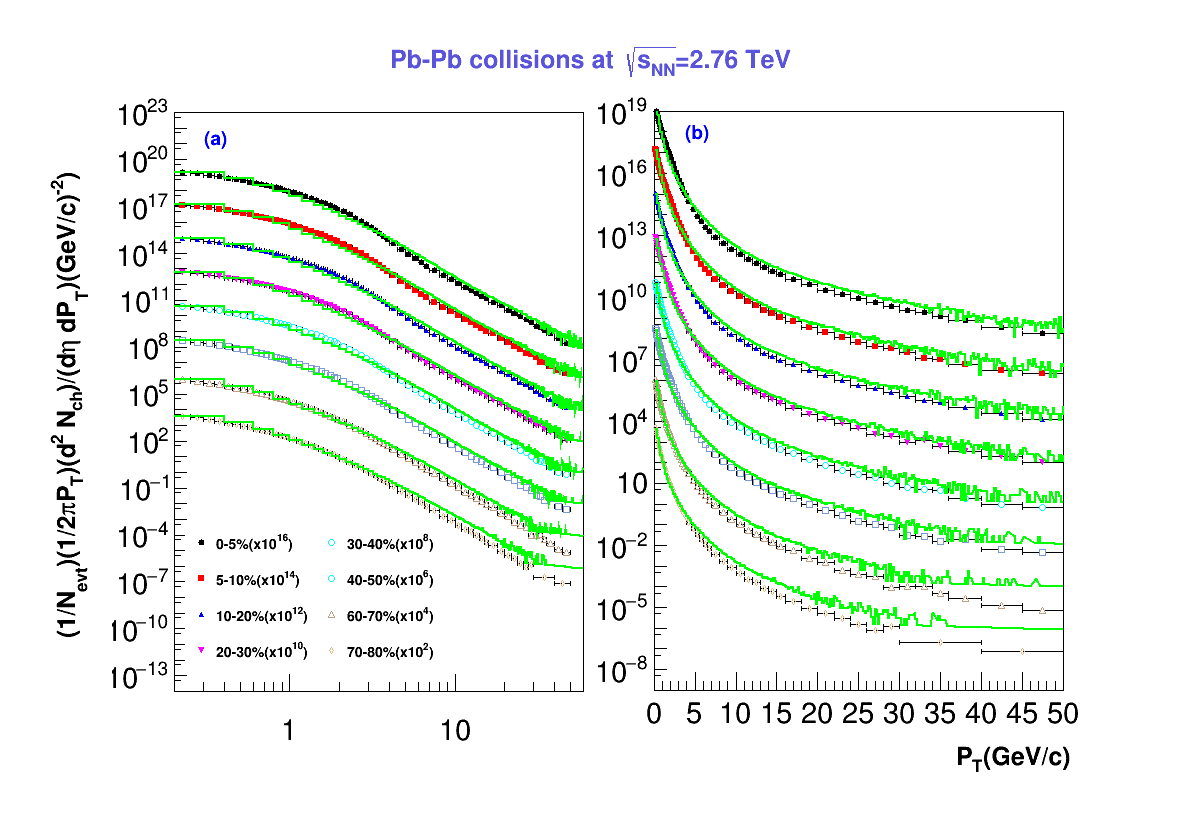}
\caption{(COLOR ONLINE)
Transverse momentum distributions of charged particles at eight centrality intervals in $Pb+Pb$ collisions at $\sqrt{s_{NN}}=2.76$ TeV, from ALICE experiment \cite{r13} (points with error bars), as compared to HIJING/Cas results (solid lines). (a) and (b) are the results in log-log and linear-log scales, respectively. For clarity, both histograms and the data are multiplied by the indicated values.}
\label{f6}
\end{center}
\end{figure}

Finally, in Fig.\ref{f6}, we compare ImHIJING and ImHIJING/Cas results with the measured transverse momentum $(p_T)$ distributions of primary charged particles at mid-pseudorapidity $(|\eta|<0.8)$ and at different centrality intervals for the reaction under study. As in $p+Pb$ case, there is almost no significance difference between the ImHIJING and ImHIJING/Cas results (not shown here). Both calculations can  describe the exponential behavior in the $p_T$ range below 6 GeV/c for all centrality intervals. However, the power law behavior at $p_T>6$ GeV/c is slightly overestimated as the centrality decreases. Notice that the $p_T$-dependence for peripheral (70-80\%) $Pb+Pb$ collisions is similar to the ALICE $p+p$ scaled reference, obtained by scaling data measured at $\sqrt{s_{NN}} =0.9$ and 7 TeV.  This again implies that final state parton re-scatterings (or jet quenching) should be included in ImHIJING(/Cas) for a better description of charged particles suppression at  $p_T>6$ GeV/c.

It should be noted that the $p_T$-spectra generated by HIJING2.0 are different from the measured ones at low and high $p_T$, with the code overestimate the data by 20\% at $ p_T<0.5$ GeV/c, and underestimate the charged yield  by a factor of about two at $p_T=1.5$  GeV/c \cite{r29}. On the other hand, the AMPT model reasonably describes the charged particles  $p_T$-spectra for $p_T< 1$ GeV/c but gives smaller values for larger $p_T$ \cite{r30}. This indicates that the recent versions of HIJING show more significant quenching of the final state hard-scattered partons than seen by LHC data.

\section{Conclusions}
Within the HIJING code that is updated with most recent parton distribution functions (ImHIJING) and a collective cascade recipe, we study the properties of lightening-like effect in $p+Pb$ and $Pb+Pb$ collisions at LHC energies. We find that the measured pseudorapidity density of charged particles as a function of collision centrality can be described by the ImHIJING if the lightening-like effect is taken into account.  The effect is shown to reduce the $dN_{ch}/d\eta_{lab}$ yield at the pseudorapidity region of the target in $p+Pb$  collisions. As for $Pb+Pb$  collisions, in addition to the reduced $dN_{ch}/d\eta$ yield at mid-pseudorapidity, the lightening-like effect builds up two peaks in the pseudorapidity region of the projectile/target $(|\eta|>8)$ as the centrality decreases.  The trend and magnitude of the transverse momentum distributions  are also in excellent agreement with the measurements at $p_T\leq6$ GeV/c for the reactions under study. However, for $p_T>6$ GeV, the measured $p_T$-distributions are  distinctively (slightly) overpredicted  for $p+Pb$ (peripheral $Pb+Pb$) collisions.

\acknowledgments

The authors would like to thank Pof. Uzhinskii for revising the code. Kh. A.-W. would like to thank the members of GEANT4 hadronic group for the hospitality and advice during his visits to CERN. The authors would also like to thank Dr Atif (umm Al-Qura university) for the Latex tips.




\begin{thebibliography}{30}
\bibitem{r1}
J. Ranft, Phys. Rev. D 37, 1842 (1988).
\bibitem{r2}
 X.$-$N. Wang and M. Gyulassy, Phys. Rev. Lett. 68 (1992) 1480;
     X. $-$N. Wang and M. Gyulassy, Phys. Rev. D 44 (1991) 3501.
 \bibitem{r3}
 W. $-$T. Deng, X. $-$N. Wang, and R. Xu, Phys. Rev. C 83 (2011) 014915;
     W. $-$T. Deng, X. $-$N. Wang, and R. Xu, Phys. Lett. B 701 (2011) 133-136;
    R. Xu, W.$-$T. Deng and X. $-$N. Wang, Phys. Rev. C 86 (2012) 051901.
 \bibitem{r4}
 K. Werner, F.$-$M. Liu, and T.Pierog, Phys. Rev. C 74 (2006) 044902.
\bibitem{r5}
 S. A. Bass et al., Prog. Part. Nucl. Phys. 41 (1998) 255.
\bibitem{r6}
 M. Mitrovski, T. Schuster, G. Graf, H. Petersen, and M. Bleicher, Phys. Rev. C 79 (2009) 044901.
\bibitem{r7}
 A. D. Martin, W. J. Stirling, R. S. Thorne, and G. Watt, Eur. Phys. J. C 63(2009) 189.
\bibitem{r8}
 Kh. Abdel-Waged and V.V. Uzhinskii, Phys. Atom. Nucl. 60 (1997) 828.
\bibitem{r9}
 Kh. Abdel-Waged and V.V. Uzhinskii, J. Phys. G: Nucl. Phys. 24 (1998) 1723.
\bibitem{r10}
B. Abelev et al. (ALICE collaboration), Phys. Rev. Lett. 110 (2013) 032301.
\bibitem{r11}
 B. Abelev et al. (ALICE collaboration), Phys. Rev. Lett. 110 (2013) 082302;
       B. Abelev et al. (ALICE collaboration), Eur. Phys. J. C 74  (2014) 3054.
 \bibitem{r12}
 E. Abbas et al. (ALICE collaboration), Phys. Lett. B 726 (2013) 610-622.
\bibitem{r13}
 B. Abelev et al. (ALICE collaboration), Phys. B 720 (2013) 52-60.
\bibitem{r14}
 T. Sjostrand and M. vanZijl, Phys. Rev. D 36 (1987) 2019.
\bibitem{r15}
 B. Andersson, G. Gustafson, and B. Nilsson-Almqvist, Nucl. Phys. B 281 (1987) 289.
\bibitem{r16}
 A. Capella, U. Sukhatme, and J. Tran Thanh Van, Z. Phys. C 3 (1979) 329.
\bibitem{r17}
 D. W. Duke and J. F. Owens, Phys. Rev. D 30 (1984) 49.
\bibitem{r18}
 M. Gluck, E. Reya, and A. Vogt, Z. Phys. C 67 (1995) 433.
\bibitem{r19}
 A. Abulencia et al. (CDF-Run II collaboration), Phys. Rev. D 75 (2007) 092006.
\bibitem{r20}
 T. Aaltonen et al. (CDF collaboration), Phys. Rev. D 78 (2008) 052006.
\bibitem{r21}
V. M. Abazov et al. (D${\O}$ collaboration), Phys. Rev. Lett. 101 (2008) 062001.
\bibitem{r22}
 K. Abdel-Waged and N. Felemban, Phys. Rev. C 91 (2015) 034908.
\bibitem{r23}
 M. I. Adamovich et al. (EMU01 collaboration), Z. Phys. A: At. Nucl. 358 (1997) 337.
\bibitem{r24}
S. Roesler, R. Engel, and J. Ranft, arXiv:hep-ph/0012252.
\bibitem{r25}
 B. Abelev et al. (ALICE collaboration) Phys. Rev. C 88 (2013) 044909.
\bibitem{r26}
 Z. W. Lin, C. M. Ko, B. $-$A. Li, B. Zhang, S. Pal, Phys. Rev. C 72 (2005) 064901.
\bibitem{r27}
 S. Pal and M. Bleicher, Phys. Lett. B 709  (2012) 82-86.
\bibitem{r28}
 J. L. Albacete, A. Dumitru, and Y. Nara, J.  Phys. Conf. Ser. 316 (2011) 012011.
\bibitem{r29}
 G. Aad et al. (ATLAS collaboration) Phys. Lett. B 710 (2012) 363-382.
\bibitem{r30}
Jun Xu and Che Ming Ko, Phys. Rev. C 83 (2011) 034904.

\end{thebibliography}
\end{document}